\pdfoutput=1

\documentclass[final,3p,times,twocolumn]{elsarticle}
\usepackage{graphicx}
\usepackage{amssymb}
\usepackage{gensymb}
\usepackage{natbib}
\usepackage{mathtools}
\usepackage{lineno}

\begin{document}

\begin{frontmatter}


\title{Laser-generated plasmas in length scales relevant for thin film growth and processing: simulation and experiment}



\author[label1]{S. B. Harris \corref{cor1}}
\ead{sumner@uab.edu}

\author[label1]{J. H. Paiste}

\author[label1]{T. J. Holdsworth}

\author[label2]{R. R. Arslanbekov}

\author[label1]{R. P. Camata}

\address[label1]{Department of Physics, University of Alabama at Birmingham, Birmingham, Alabama 35294, USA}
\address[label2]{CFD Research Corporation, Huntsville, Alabama, 35806, USA}
\cortext[cor1]{Corresponding author}

\begin{abstract}
In pulsed laser deposition, thin film growth is mediated by a laser-generated plasma, whose properties are critical for controlling the film microstructure. The advent of 2D materials has renewed the interest in how this ablation plasma can be used to manipulate the growth and processing of atomically thin systems. For such purpose, a quantitative understanding of the density, charge state, and kinetic energy of plasma constituents is needed at the location where they contribute to materials processes. Here we study laser-induced plasmas over expansion distances of several centimeters from the ablation target, which is the relevant length scale for materials growth and modification. The study is enabled by a fast implementation of a laser ablation/plasma expansion model using an adaptive Cartesian mesh solver. Simulation outcomes for KrF excimer laser ablation of Cu are compared with Langmuir probe and optical emission spectroscopy measurements.  Simulation predictions for the plasma-shielding threshold, the ionization state of species in the plasma, and the kinetic energy of ions, are in good correspondence with experimental data. For laser fluences of 1-4 J/cm$^2$, the plume is dominated by Cu$^0$, with small concentrations of Cu$^{+}$ and electrons at the expansion front. Higher laser fluences (e.g., 7 J/cm$^2$) lead to a Cu$^{+}$-rich plasma, with a fully ionized leading edge where Cu$^{2+}$ is the  dominant species. In both regimes, simulations indicate the presence of a low-density, high-temperature plasma expansion front with a high degree of ionization that may play a significant role in doping, annealing, and kinetically-driven phase transformations in 2D materials.

\end{abstract}

\journal{Journal of Physics D: Applied Physics}
\end{frontmatter}
\section{Introduction}
\label{S:1}
Low-temperature plasmas (LTPs) have long been recognized as effective in enhancing physical and chemical processes that take place during materials synthesis \cite{Chang2017,Allain2017,Oehrlein2018}. The emergence of two-dimensional (2D) materials has renewed the interest in using LTPs for engineering these systems at the atomic scale. Well-known and as-of-yet unexplored LTP regimes may offer nonequilibrium and reactive environments that are not accessible to other 2D materials fabrication approaches. Recent examples include the engineering of edge sites of WS$_2$ in plasma-enhanced atomic layer deposition \cite{Balasubramanyam2019a} and the formation of Ohmic contacts to PdSe$_2$ by exposure to an argon plasma \cite{Oyedele2019}.

LTPs generated by laser ablation are of particular interest because of their rich chemistry and spatiotemporal phenomena. The chemical diversity, variety of gas backgrounds, and shockwave characteristics of the laser plume are highly conducive to kinetic control of materials synthesis \cite{Harilal2016}. Laser-induced LTPs also exhibit wide tunability of plasma parameters. This includes wavelength dependence of the electron energy distribution function \cite{Liu2019} and strong gradients of density and temperature that cause deviations from local thermodynamic equilibrium \cite{Liu2016}. This suite of attractive characteristics has long been considered significant in pulsed laser deposition (PLD) of thin films \cite{Geohegan1992a}. Measurements of laser plasma parameters using Langmuir probes, time-of-flight mass spectroscopy, optical emission spectroscopy, laser-induced fluorescence, as well as photography and imaging techniques have established numerous correlations between plasma parameters and thin film characteristics \cite{Willmott2000}. Increasing the kinetic energy of plasma species, for example, leads to increased depth of implantation and higher density of defects in the films. Conversely, low kinetic energy deposition creates film growth conditions near thermodynamic equilibrium \cite{Geohegan1994}. These and many other correlations represent a substantial body of knowledge but do not provide an exhaustive picture of the plasma environment during thin film growth by PLD. This is due in part to the difficulty in gathering direct experimental information on the strongly time and space dependent plasma parameters at the location of film growth. Moreover, growth and modification of 2D materials has raised new questions about the ion densities, kinetic energies, and plasma temperatures that are conducive to manipulation of these atomically thin systems. In this context, laser plasma simulations may offer insights into the plasma characteristics in the vicinity of the processing area which, in conjunction with experiments, could reveal plasma conditions for controlling processes such as doping, defect annealing, and kinetically-driven phase transformations.

Although simulations of laser-induced plasmas have been carried out for decades, the vast majority of reported results concentrate on laser plume length scales of up to tens of microns. Some studies extend the plume behavior up to a few millimeters. These length scales are most useful for studying processes that take place in close proximity to the ablation target, such as laser micromachining \cite{Palya2018}, target elemental analysis \cite{Shabanov2014}, and target processing \cite{Cadot2018}. For thin film deposition and 2D materials processing, however, it is desirable to perform simulations out to significantly longer distances, since substrates are typically placed 4-10 cm away from the target. Simulating the laser plasma out to these length scales has been a challenge because of the long calculation times associated with currently available codes and computing platforms. As a result, very few computational investigations for similarly long  distances are found in the literature \cite{Pathak2008,Sizyuk2014,Sharad2017}. In this paper, we implement a previously described laser ablation model employing a solver with an adaptive Cartesian mesh (ACM). The vastly improved efficiency and time resolution of this implementation allows us to simulate the laser-produced plasma expansion out to the realistic PLD distances of centimeters. This includes simulation of leading edge characteristics of the plume that may meaningfully be fed into mesoscopic models of plasma-surface interaction that ultimately govern thin film growth and modification. We compare the simulation predictions with Langmuir probe and optical emission spectroscopy measurements of the plasma generated during ablation of copper (Cu). These comparisons help clarify the conditions under which our plasma simulations are suitable for describing laser-generated LTPs and where opportunities exist for expanding their domain of applicability.

\section{Model Description}
\label{S:2}
The simulation of laser-induced plasmas is a mature field with a variety of kinetic \cite{Pietanza2010}, fluid \cite{Chen2005}, and hybrid models \cite{Palya2018} described in the literature. The basic simulation of material ablation by a pulsed laser requires the integration of processes of ejection of ablated material upon laser irradiation with processes of plasma plume formation and expansion. These two stages are of course coupled, with significant mutual interference. Processes of material removal involve complicated light-matter interactions whose physics depends strongly on the laser wavelength, pulse duration, and target material properties. Thermal evaporation, subsurface boiling, supercritical boiling, and hydrodynamic sputtering vary in their relative importance in multiple regimes \cite{KellyandMiotello}. For low laser fluence ($\sim$ 1 J/cm$^2$), thermal evaporation is dominant, whereas additional mechanisms need to be considered in the high fluence regime. If temperatures approach or exceed the critical temperature of the target material, for example, supercritical processes take place. This has been accounted for in robust recent models that include both, surface and volume removal mechanisms \cite{Autrique2013}. Once the material has been ejected, models of plasma formation and expansion are needed. Numerous collisional and radiative processes, including inverse Bremsstrahlung, electron impact ionization, and electron impact excitation/deexcitation can be accounted for to determine the electron and ion populations \cite{Autrique2013a}. The  evolution of the plasma can be calculated using kinetic or fluid models. Despite their accuracy and fundamental appeal, kinetic models remain impractical for simulations out to several centimeters and microsecond timescales \cite{Palya2019}. Fluid models employing the assumption of local thermodynamic equilibrium (LTE) are the most popular \cite{Wang2016}, while two-temperature models have also been evaluated \cite{Autrique2013a, Oumeziane2016}.

For our purpose of developing laser plume simulations out to long distances, we have opted for a simple model that couples laser-induced surface evaporation with fluid plasma expansion \cite{Chen2005}. Despite its simplifying assumptions about collisional/radiative processes in the plasma, and the exclusion of multiphase processes of target material removal, the model is sufficiently sophisticated to be applicable to realistic simulations under a limited set of conditions. LTE is assumed throughout the simulation, including the initial stages of plasma formation. The model neglects material removal by phase explosion. In general, phase explosion contributes significantly to the ablation process at high laser powers and it is commonly accepted that deviation from LTE occurs in very early stages of plasma formation \citep{Autrique2013,Autrique2013a}. Our main focus is on PLD and 2D materials processing at relatively low laser fluence of 1-4 J/cm$^{2}$ (0.5-2$\times$10$^{8}$ W/cm$^{2}$) for a 248 nm, 20 ns laser pulse. Under these conditions, which are routinely used to grow high-quality, epitaxial films in high vacuum, phase explosion is not expected to contribute significantly to material removal \cite{Autrique2013}. The expanding gas in this regime contains a low density of ionized species in comparison to high fluence cases ($>$10 J/cm$^{2}$).

In what follows we summarize the one-dimensional (1D) version of the chosen model. The 1D description effectively entails a flat plasma expansion front. This is a suitable geometry at distances from the target that are smaller than the laser spot size. The 1D version provides therefore a good approximation during the initial stages of the expansion (i.e., several hundreds of ns after the laser pulse). It is thus reasonable to expect quantitative accuracy from the model for the early expansion. Three-dimensional (3D) effects will become increasingly important as the expansion proceeds to distances that are larger than the spot size. Accordingly, the focus of the analysis at longer distances from the target should be on qualitative trends rather than on quantitative predictions. The essential features of the model have been described elsewhere \cite{Chen2005}, while its numerical implementation carried out here is entirely novel. The simulation is implemented in a multi-dimensional setting in a state-of-the-art, open-source ACM framework \cite{basilisk.fr}. Existing compressible solvers available in this framework (Riemann and so-called all-Mach solvers \cite{basilisk.fr}) have been adapted to include a complex equation-of-state (EoS) combined with the Saha equilibrium model describing the state of a multi-species laser-generated plasma, comprising ground-state metal species, several metal ion species, as well as electrons. The use of the dynamic ACM approach is essential since the initial ($t <$ 20 ns) spatial resolution required near the target is tenths of microns, while the simulation extends up to several centimeters. As the plasma plume expands ($t >$ 20-50 ns), the fine near-target resolution is no longer required for most of the simulation domain, allowing the grid to be coarsened. Proper resolution of the moving plasma front is still necessary, however, throughout the entire plasma dynamics until the front reaches the substrate (several centimeters away from the target). Only the very narrow plasma front region needs to be resolved for this purpose, while the remainder of the computational mesh can stay coarse. The ACM technique thus makes the computation fast over the large distances from the target ($>$ 1-5 cm) and over long times ($t > 1$ $\mu$s). This is already noticeable in the 1D results reported here and is even more important for multi-dimensional implementations. The details of the compressible multi-dimensional calculations, with the complex EoS and its numerical scheme, will be described elsewhere. 

\subsection{Target heating, melting and evaporation}
The interaction of the laser pulse with the target is assumed to follow Beer's Law with a constant and uniform absorption coefficient $\alpha$ and reflectivity $\mathcal{R}$ (Eq. \ref{beer}). This absorption results in a heat source term in the heat diffusion equation (Eq. \ref{heat}), that is solved to account for the heating of the target, whose laser-exposed, thin top layer eventually melts and vaporizes. 
\begin{equation}\label{beer}
    I(x,t) = I_{0} e^{-\alpha x} [1-\mathcal{R}]
\end{equation}

\begin{equation}\label{heat}
    \frac{\partial T(x,t)}{\partial t} = \frac{\partial}{\partial x}\bigg[\frac{\kappa}{(C_{p}\rho_m)} \frac{\partial T(x,t)}{\partial x}\bigg] + \frac{\alpha}{C_{p}\rho_m} I(x,t)
\end{equation}
In Eq. \ref{heat}, $\kappa$, $C_{p}$, and $\rho_m$ represent the thermal conductivity, heat capacity, and mass density of the target material, respectively. 
The pressure of the vapor that forms on the surface of the target is calculated from the Clausius-Clayperon equation (Eq. \ref{claus}), where $\Delta H_{lv}$ is the heat of vaporization, $T_{s}$ and $T_{b}$ are the surface and boiling temperatures at 1 atm ($p_{0}$), and $R$ is the ideal gas constant.
\begin{equation}\label{claus}
    p_{vap}(T_{s}) = p_{0} \exp\bigg[\frac{\Delta H_{lv} (T_{s} - T_{b})}{RT_{s}T_{b}} \bigg] 
\end{equation}

\begin{equation}\label{idealgas}
    \rho_{vap,s} = \frac{p_{vap}}{kT_{s}}
\end{equation}
The vapor density at the surface is calculated from the ideal gas law (Eq. \ref{idealgas}), which is used to approximate the average normal velocity component (Eq. \ref{velocity}) by assuming a Maxwellian velocity distribution.
\begin{equation}\label{velocity}
    \nu_{vap,s} = \sqrt{\frac{2kT_{s}}{\pi m}}
\end{equation}

\subsection{Expansion of the plume}
The expansion of the vapor into vacuum is described by the equations of fluid mechanics, assuming conservation of mass (Eq. \ref{massdensity}), momentum (Eq. \ref{momentum}), and energy (Eq. \ref{cons_energy}),

\begin{equation}\label{massdensity}
    \frac{\partial \rho}{\partial t} = -\frac{\partial \rho \nu}{\partial x}
\end{equation}
    
\begin{equation}\label{momentum}
    \frac{\partial \rho \nu}{\partial t} = -\frac{\partial}{\partial x}[p + \rho \nu^{2}] 
\end{equation}

\begin{equation}\label{cons_energy}
\begin{multlined}
    \frac{\partial}{\partial t} \bigg[\rho \bigg(E + \frac{\nu^2}{2} \bigg) \bigg] = - \frac{\partial}{\partial x} \bigg[\rho \nu \bigg(E + \frac{p}{\rho} + \frac{\nu^2}{2} \bigg)  \bigg]\\
    + \alpha_{IB} I_{laser} - \epsilon_{rad}
\end{multlined}
\end{equation}
where $\rho$, $\rho \nu$, and $\rho E$ represent the plasma mass density, momentum density, and internal energy density, respectively. 

In Eq. \ref{cons_energy}, $\alpha_{IB}$ is the absorption coefficient for inverse Bremsstrahlung and $I_{laser}$ is the laser irradiance. Their product acts as an energy source for the fluid, while $\epsilon_{rad}$ is the emitted energy from the Bremsstrahlung process, playing the role of an energy sink. Assuming a Maxwellian velocity distribution for the electrons, $\epsilon_{rad}$ is given by Eq. \ref{radiation} \cite{Spitzer2006}, where $n_e$, $n_{i1}$, and $n_{i2}$ are the number densities of electrons, Cu$^{+}$, and Cu$^{2+}$, respectively, with $Z_1$ = 1 and $Z_2$ = 2.
\begin{equation}\label{radiation}
    \epsilon_{rad} = \bigg( \frac{2 \pi k T}{3 m_e}\bigg)^{1/2} \frac{32 \pi e^6}{3 h m_e c^3} n_e (Z^2_{1} n_{i1} + Z^2_{2} n_{i2})
\end{equation}

The local pressure $p$ and the temperature $T$ of the expanding vapor are related to $\rho$ and $\rho E$ by \cite{Zeldovich2002} 

\begin{equation}\label{idealgas2}
    p = (1 + x_{e}) \frac{\rho k T}{m}
\end{equation}

\begin{equation}\label{internalenergy}
    \rho E = \frac{\rho}{m}\bigg[ \frac{3}{2}(1 + x_{e}) k T + IP_1 x_{i1} + (IP_1 + IP_2) x_{i2} \bigg]
\end{equation}
where $x_{e}$ is the degree of ionization of the vapor, and $x_{i1}$ and $x_{i2}$ represent the fractions of singly and doubly ionized atoms (Cu$^{+}$ and Cu$^{2+}$, in this case). $IP_1$ and $IP_2$ denote the first and second ionization potentials of Cu.

\subsection{Plasma formation}
The ionization of the vapor, and consequent plasma formation, is governed by a set of Saha equations that take into account the concentrations of Cu$^{0}$, Cu$^{+}$, and Cu$^{2+}$:
\begin{equation}\label{Saha_1}
    \frac{x_e x_{i1}}{x_0} = \frac{1}{n_{vap}} \bigg( \frac{2 \pi m_e k T}{h^2}\bigg)^{3/2} exp\bigg( -\frac{IP_1}{kT} \bigg)
\end{equation}
\begin{equation}\label{Saha_2}
    \frac{x_e x_{i2}}{x_{i1}} = \frac{1}{n_{vap}} \bigg( \frac{2 \pi m_e k T}{h^2}\bigg)^{3/2} exp\bigg( -\frac{IP_2}{kT} \bigg)
\end{equation}
The number density of the vapor $n_{vap}$ is obtained from the vapor density and the mass of Cu. The degree of ionization and fractions of Cu$^{0}$, Cu$^{+}$, and Cu$^{2+}$ species are again represented by $x_e$, $x_0$, $x_{i1}$, and $x_{i2}$, respectively. 

Eqs. \ref{idealgas2} through \ref{Saha_2} represent a so-called non-ideal-gas EoS, $p=p(\rho,E)$ which has been implemented in our ACM framework using Newton iteration schemes and numerical differentiation (e.g., the speed of sound required in the compressible schemes is computed as $c^2 = \partial{p}/\partial{\rho}\big|_s$, where $S$ is the enthalpy).

The vacuum background is characterized by a fixed number density of neutral species, which can be adjusted to match experimental conditions of interest. Conservation of matter and charge are accounted for by using Eqs. \ref{matter} and \ref{charge}.
\begin{equation}\label{matter}
    x_0 + x_{i1} + x_{i2} = 1
\end{equation}
\begin{equation}\label{charge}
    x_{i1} + 2 x_{i2} = x_{e}
\end{equation}


\subsection{Laser absorption by the plasma}
Once the plasma has formed in front of the target, some of the laser pulse is absorbed via inverse Bremsstrahlung in which a free electron absorbs a photon near a heavy particle. The absorption coefficients for electron-neutral and electron-ion inverse Bremsstrahlung are given by Eqs. \ref{e-nIB} and \ref{e-iIB}, respectively \cite{Rakziemski1989}.
\begin{equation}\label{e-nIB}
    \alpha_{IB, e - n} = \bigg[ 1- \exp\bigg( -\frac{h c}{\lambda k T}\bigg)\bigg] Q n_e n_0
\end{equation}
\begin{equation}\label{e-iIB}
\begin{multlined}
    \alpha_{IB, e - i} = \bigg[ 1- \exp\bigg( -\frac{h c}{\lambda k T}\bigg)\bigg] \frac{4 e^6 \lambda^3 n_e}{3 h c^4 m_e}\\
    \times \bigg( \frac{2 \pi}{3 m_e k T}\bigg)^{1/2} (Z^2_1 n_{i1} + Z^2_2 n_{i2})
\end{multlined}
\end{equation}
where $Q$ is the cross section for photon absorption by an electron during a collision with neutral species (taken to be 10$^{-36}$ cm$^2$) \cite{Rakziemski1989}.

\section{Experimental Details}
\label{S:3}
\begin{figure}[h]
    \centering
    \includegraphics[width=\linewidth]{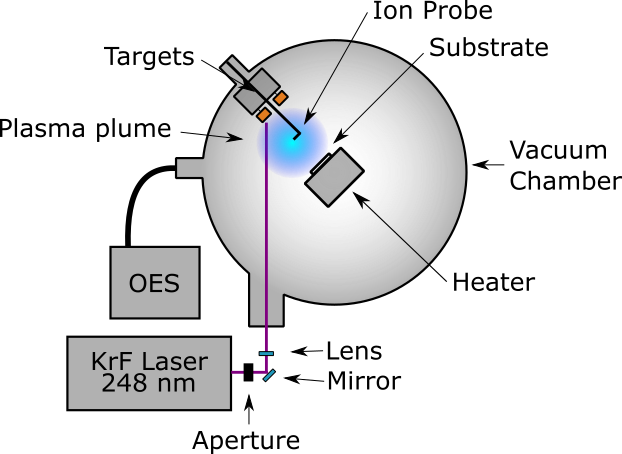}
    \caption{Schematic of the pulsed laser deposition system, including Langmuir probe and optical emission spectroscopy (OES) diagnostics. A 10 $\times$ 10 mm planar probe is used to acquire data from the ions in the plasma plume, whereas electrons are studied with a 100-$\mu$m diameter cylindrical probe of 4.5 mm length. Each probe is attached to a micrometric linear translation stage that allows its placement at distances between zero and $5.500 \pm 0.005$ cm from the target.}
    \label{experiment_schematic}
\end{figure}

The properties of a laser-generated plasma corresponding to the simulation were measured with Langmuir probes of two different geometries in the PLD system shown schematically in Fig. \ref{experiment_schematic}. A 10 $\times$ 10 mm square planar probe was used for measurements of the density and the time of flight (TOF) of ions. Under strong negative bias, only positive, ionic species are collected by the planar probe. The ion current measured as a function of the time delay from the laser pulse can be used to deduce how the ion density of the plasma changes with time at the position of the probe \cite{Doggett2009}. For measuring electron properties, a cylindrical probe consisting of a 100-$\mu$m diameter, 4.5-mm long platinum wire was employed. Under positive bias, cylindrical probe theories permit extraction of the electron temperature ($T_e$) from the current vs. bias voltage characteristics ($I$-$V_{Bias}$) \cite{Weaver1999}. Each probe was placed along the direction normal to the surface of the Cu ablation target and was attached to a linear translation vacuum feedthrough, in order to be positioned at distances between zero and $5.500 \pm 0.005$ cm from the target. The probes are connected to a bipolar bias voltage source, and the collected current is determined from the voltage drop across a 10 $\Omega$ resistor, connected to ground.
Curves of probe current vs. bias voltage can be measured by stepping the bias from negative to positive and collecting current pulses at every bias value. Averaging of numerous current pulses yields well-defined current traces, peak currents, and values of the electron temperature $T_e$. Although the exact number of pulses used in each probe measurement varied over the course of experiments, all probe data reported here involve averaging of a minimum of 255 pulses. The Cu target was well-polished before each experiment to aid in reproducibility and in matching the conditions of the simulation. The pressure of the chamber was kept at $3.0\times10^{-5}$ torr during ablation experiments with a KrF (248 nm) Lambda Physik LPX 305i excimer laser, focused to spot on the target with an area of 0.0439 cm$^2$.

Optical emission spectroscopy (OES) data from the PLD plume was acquired with an Acton Research Corporation SpectraPro 500i, 0.5 meter spectrometer using a 1200 grooves/mm grating blazed at 300 nm. Each OES spectrum was collected with a quartz, optical fiber aimed at the substrate mounting plate in the PLD chamber. 

The simulation parameters were set to be consistent with our PLD environment, with the obvious caveat that this is a 1D implementation of the model, whereas the experiment takes place in three dimensions. Accordingly, comparisons between experiments and simulation predictions at long distances from the target are primarily qualitative. The laser wavelength was set to 248 nm and a top-hat intensity temporal profile was used with a FWHM of 20 ns. Laser power densities of 1-5$\times10^{8}$ W/cm$^{2}$ were used, which correspond to laser fluences of 2-10 J/cm$^{2}$. The number density of atoms in the  vacuum background of the simulation was set to 10$^{18}$ m$^{-3}$, to match the $3.0\times10^{-5}$ torr vacuum environment where probe and OES measurements were carried out.

\section{Results and Discussion}
\label{S:4}
\begin{figure}[t]
    \centering
    \includegraphics[width=\linewidth]{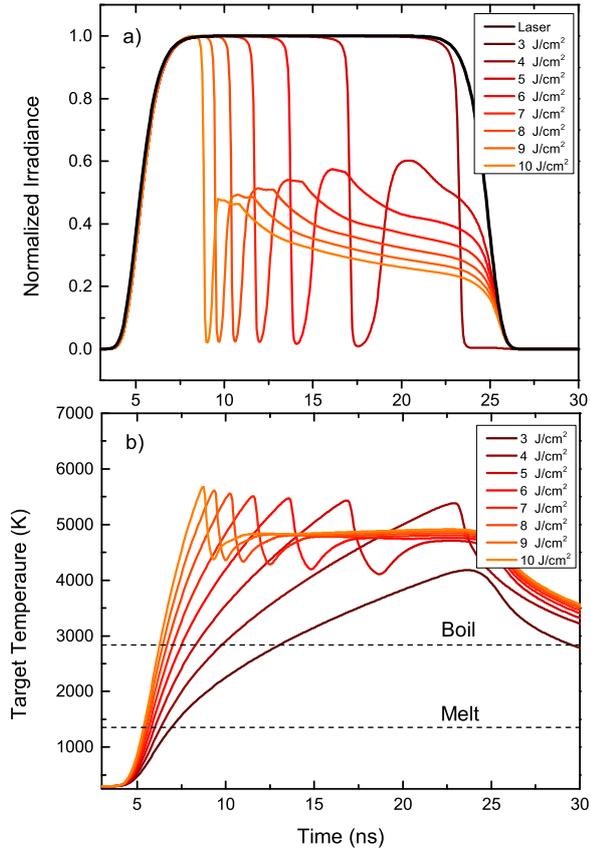}
    \caption{a) Simulated target irradiance for different values of laser fluence using a 248 nm, 20 ns laser pulse with a top-hat temporal profile. As the laser fluence increases, the sharp decrease of target irradiance, indicating onset of plasma shielding, occurs at earlier times. b) Simulated target temperature at different laser fluences shows the peak temperature is reached right before the onset of plasma shielding. The peak temperature increases with increasing laser fluence.}
    \label{target_absorp_temp}
\end{figure}

 We first examine some general characteristics of the simulation in light of ion probe and OES measurements. Fig. \ref{target_absorp_temp}a shows the effect of laser fluence on the irradiance calculated to reach the target. The temporal profile of the laser pulse is also shown for reference. The corresponding target heating curves are seen in Fig. \ref{target_absorp_temp}b. A sharp decrease in the target irradiance signals the onset of shielding of the laser beam by the plasma. The peak target temperature occurs right before the onset of plasma shielding. For fluences of 3 J/cm$^2$ and below, the target absorbs 100\% of the laser energy, which results in a smooth heating curve with the maximum temperature of about 4180 K at $t\approx 23$ ns, coinciding with the end of the laser pulse. At greater fluence values, the model predicts the onset of shielding to occur at earlier times during the laser pulse. In each case, the drop in target irradiance is accompanied by a sharp decline in target temperature at first, followed by a reduced heating rate for the remainder of the laser pulse. An important quantity to watch in these calculations is the peak target temperature. As observed in Fig. \ref{target_absorp_temp}b, the solution of the heat diffusion equation predicts peak target temperatures that do not exceed 5700 K for the range of laser fluences investigated (2-10 J/cm$^{2}$ with $\lambda$ = 248 nm and 20 ns pulse duration). This is significantly below the critical temperature of Cu, $T_{crit}$ = 7800 K, indicating that supercritical processes do not play a significant role in the target ablation for these conditions \cite{Autrique2013}. For the limited scope of the present investigation, the Cu target reflectivity has been fixed at $\mathcal{R}$=0.34 \cite{Mao1997}. This is of course a simplistic assumption of the model, since the Cu reflectivity must vary as it passes through the solid-liquid phase transition. However, a parametric study of the effect of $\mathcal{R}$ on the target heating curves reveals that this admittedly oversimplified scenario does not alter the conclusion that supercritical processes can be safely discounted under our conditions. Specifically, simulations using a broad range of meaningful values of reflectivity ($\mathcal{R} = 0.14-0.95$) yield a maximum peak temperature of $\sim$ 6200 K, which is still below $T_{crit}$. 

\begin{figure}[t]
    \centering
    \includegraphics[width=\linewidth]{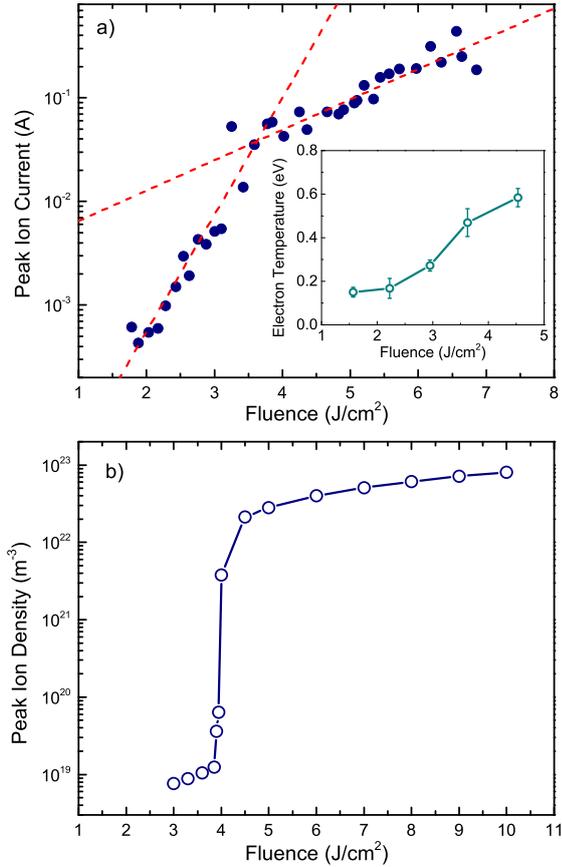}
    \caption{a) Peak ion current collected by the planar probe at 5.5 cm away from the target, with a bias voltage of $-$50 V. The ion current grows exponentially with the laser fluence. The change in the rate of exponential growth noted at $\sim$ 3.6 J/cm$^2$ corresponds to the threshold of plasma absorption of the laser pulse. The inset displays the electron temperature measured with the cylindrical probe for fluences near the plasma shielding  threshold. b) Simulated Cu$^{+}$ density vs. fluence at 2.0 cm from the target, showing the occurrence of the plasma-shielding transition at a laser fluence of $\sim$ 3.9 J/cm$^2$.}
    \label{peak_dens_vs_fluence}
\end{figure}

Plasma shielding has a nonlinear impact on the density of the laser-generated plume. This effect can be detected by Langmuir probes and offers a first opportunity for comparing the simulation results with experiments. Fig. \ref{peak_dens_vs_fluence}a shows the peak current measured by the planar ion probe as a function of laser fluence. The dependence of the ion current with the fluence clearly reveals two distinct regimes. A change in slope in the semi-log plot of the peak ion current vs. fluence is observed at $\sim$ 3.6 J/cm$^2$. This ``kink'' in the ion current behavior has been observed in the ablation of other materials, and it has been attributed to the threshold for laser absorption by the plasma \cite{Geohegan1994}. Comparing the experimental value of $\sim$ 3.6 J/cm$^2$ to the target irradiance curves in Fig. \ref{target_absorp_temp}a, it is apparent that the change in slope of the experimentally collected ion current is consistent with entry into the plasma shielding regime, predicted by the simulations to set in around 3-4 J/cm$^2$ of laser fluence. It is interesting to note that no luminous plume was visible in the PLD chamber for fluences below 2.3 J/cm$^2$. A very weak plume becomes discernible to the eye at approximately 2.6 J/cm$^2$. Plumes with strong optical emission are observed for  fluences $\sim$ 3 J/cm$^2$. The inset of Fig. \ref{peak_dens_vs_fluence}a shows the electron temperature measured with the cylindrical probe for fluences near the plasma-shielding threshold. The increasing trend of the electron temperature is in agreement with the regime transition noted in the ion current data. 
\begin{figure*}[t]
    \centering
    \includegraphics[width=\textwidth]{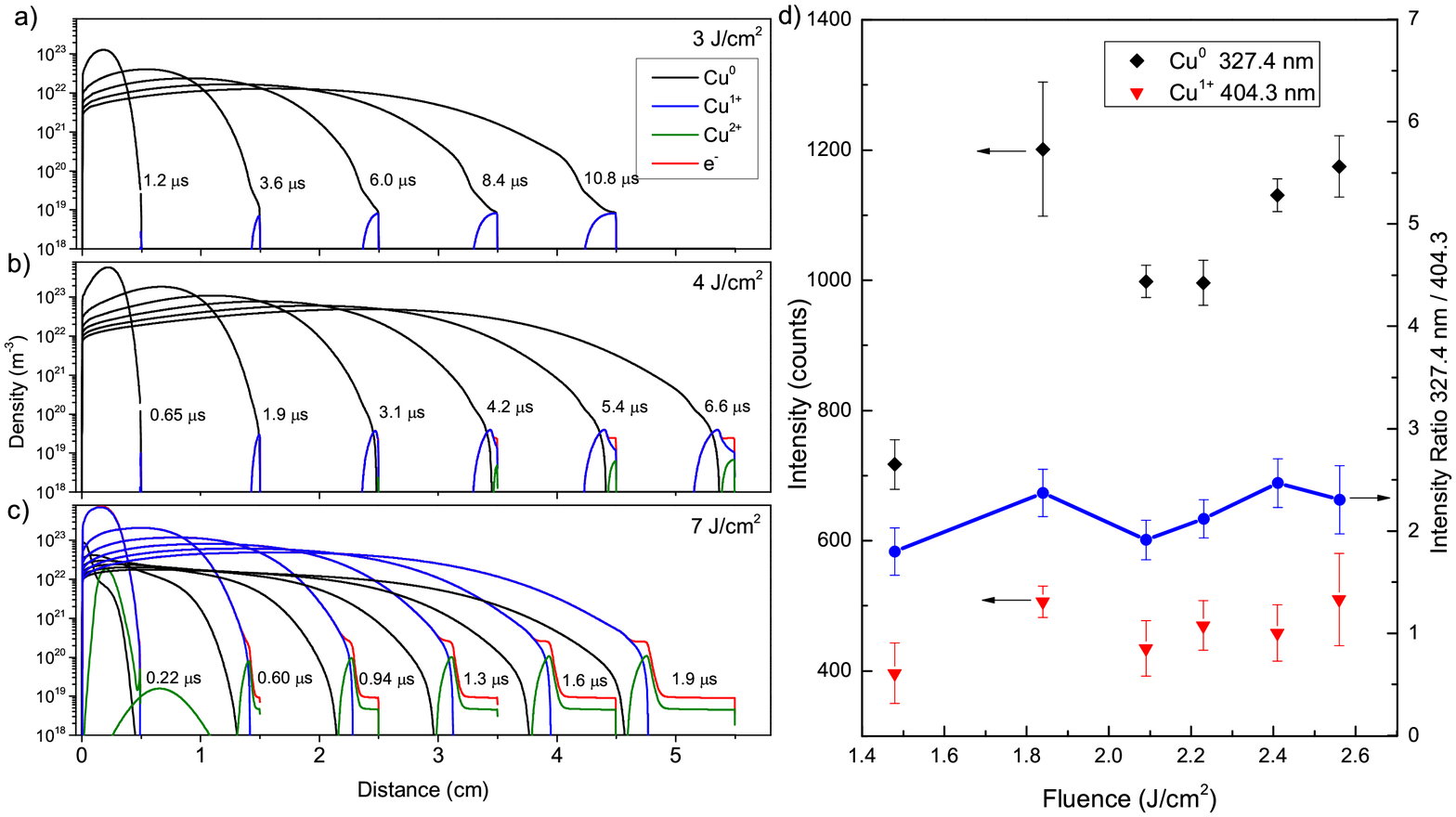}
    \caption{Density of e$^{-}$, Cu$^{0}$, Cu$^{+}$, and Cu$^{2+}$ over the 5.5-cm expansion length from the location of the PLD target for laser fluence of a) 3.0 J/cm$^{2}$, b) 4.0 J/cm$^{2}$, and c) 7.0 J/cm$^{2}$. For low fluence, the plume is dominated by Cu$^{0}$ with a leading edge containing Cu$^{+}$ and electrons. Moderate fluence shows the appearance of Cu$^{2+}$ species and high fluence leads to a highly ionized plume dominated by Cu$^{+}$ species. The expansion front is fully ionized with doubly ionized species arriving at the growth substrate first. The time lag between deposition of Cu$^{2+}$ and Cu$^{+}$ becomes more pronounced the greater the expansion length. d) Comparison of optical emission spectrum data for the Cu PLD plume shows constant ratio of Cu$^0$ to Cu$^{+}$ intensities for the 1.5-2.6 J/cm$^2$ fluence range, with no detectable Cu$^{2+}$, in general agreement with simulations.} 
    \label{expansion_plots}
\end{figure*}

The simulated peak ion density also shows a clear, abrupt shift between low- and high-density regimes at a laser fluence of $\sim$ 3.9 J/cm$^2$ that can be ascribed to the plasma-shielding transition. This is seen in Fig. \ref{peak_dens_vs_fluence}b, which displays the simulated peak density of Cu$^{+}$ ions at 2.0 cm from the target. At low laser fluence, a modest gradual increase in ion density is predicted, varying upward from $7.7\times10^{18}$ at 3.0 J/cm$^2$ to $1.8\times10^{19}$ m$^{-3}$ at $\sim$ 3.9 J/cm$^2$. The target absorption profiles in this range of fluences (Fig. \ref{target_absorp_temp}a) indicate that virtually all the energy from the laser pulse is delivered to the target in this regime, with peak target temperatures remaining relatively low ($<$ 4180 K). In the vicinity of 3.9 J/cm$^{2}$, a highly nonlinear response is observed in the simulated ion density for small changes in the laser fluence. This nonlinear response clearly corresponds to a transition to the plasma-shielding regime. The steep increase in the number of ions results from the heating of the plasma due to strong absorption of the laser energy via the included mechanism of inverse Bremsstrahlung. Further increases in fluence beyond the transition cause again a modest, steady increase in ion density, now governed by the plasma shielding regime. 

Although the simulation captures the transition between vapor transparency and plasma shielding, the predicted climb in ion density with increasing laser fluence appears significantly steeper than the one observed in the experiment. This overestimate is likely due (at least in part) to our simplifying assumptions regarding collisional and radiative processes in the plasma. We recall that only Bremsstrahlung (Eq. \ref{radiation}) and inverse Bremsstrahlung (Eqs. \ref{e-nIB} and \ref{e-iIB}) are included in the model. The addition of other processes such as photoionization/recombination, impact excitation/de-excitation, and radiative decay \cite{Autrique2013a}, may lead to a more accurate slope in the ion density transition. It is also expected that inclusion of multi-dimensional effects (e.g., 2D-axisymmetric simulation) may affect this behavior. Evidently, inclusion of these additional processes and higher dimensionality will come at a computational cost.

\begin{figure}[ht!]
    \centering
    \includegraphics[width=\linewidth]{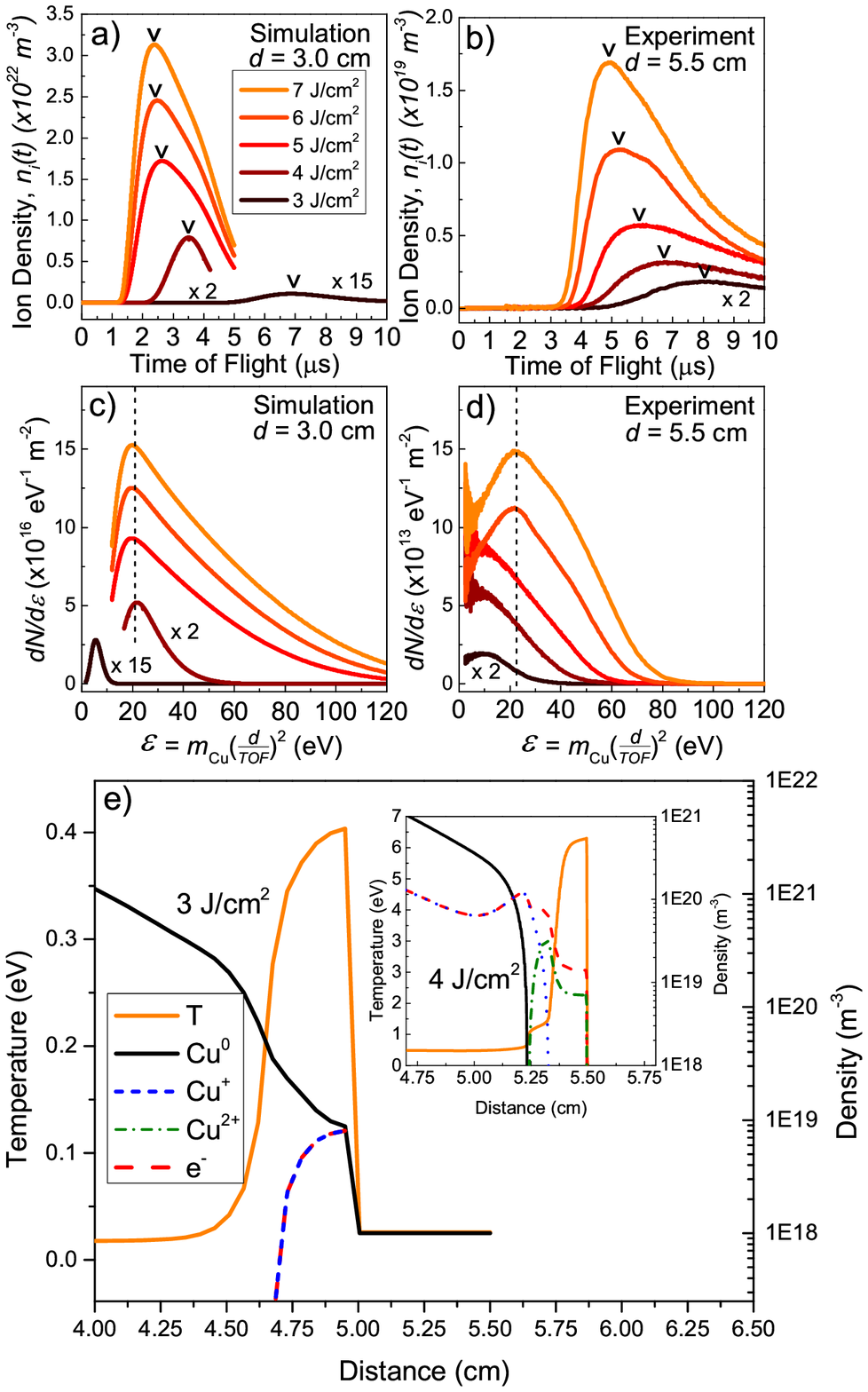}
    \caption{a) Simulated and b) measured ion densities at distance $d$ from the target, as a function of ion time of flight (TOF). Experimental ion densities measured by the planar Langmuir probe with a bias of $-$50 V. c) Simulated and d) measured $dN/d\varepsilon$ distributions obtained from the $n_i(t)$ vs. TOF curves by converting TOF to $\varepsilon = \frac{1}{2} m_{\mathrm{Cu}} \left(\frac{d}{\mathrm{TOF}}\right)^2$. e) Spatial profiles for temperature and density of the various plume species at long distances from the target. The main panel shows profiles for a laser fluence below the plasma shielding threshold (3 J/cm$^2$), while the curves in the inset correspond to a fluence just above the shielding threshold (4 J/cm$^2$). A well-defined temperature peak develops at the leading edge of the plume as a result of fluid expansion against the 10$^{18}$ atoms/m$^{3}$ vacuum background. The intensity of this temperature peak scales with the laser fluence.} 
    \label{leading_edge}
\end{figure}

Following its formation and heating, the plasma will expand outwards into the vacuum. Fig. \ref{expansion_plots} shows the simulated densities of e$^{-}$, Cu$^{0}$, Cu$^{+}$, and Cu$^{2+}$ for three laser fluence values out to a typical substrate mounting distance of 5.5 cm: a) 3 J/cm$^{2}$, below plasma-shielding threshold; b) 4 J/cm$^{2}$, near plasma-shielding threshold; and c) 7 J/cm$^{2}$, in the strong plasma shielding regime. Below the plasma-shielding threshold, the simulated plume is primarily composed of Cu$^0$ species with low densities of electrons and Cu$^{+}$ concentrated on the expansion front. The translational kinetic energy of species in the plasma front is relatively low ($\sim 6$ eV), taking 10.8 $\mu$s for them to reach 4.5 cm. For a laser fluence near the plasma-shielding threshold (4 J/cm$^{2}$), the electrons and ionic species are similarly confined to the front but with the gradual appearance of Cu$^{2+}$ as the expansion progresses. This is due to the formation of an increasingly hot leading edge of the plume. Noteworthy changes are seen in the composition of the plume at high laser fluence (7 J/cm$^{2}$; strong plasma shielding). Cu$^{+}$ becomes the dominant species and Cu$^{2+}$ is now observed at much earlier stages of the expansion. Perhaps most significantly, the leading edge of the plume achieves complete ionization.

In an attempt to experimentally assess the concentration of ions and neutrals in the Cu plume, we carried out OES measurements as described in the experimental section. The plume has strong Cu$^0$ lines and low-intensity Cu$^{+}$ lines for laser fluences in the 1.5-2.7 J/cm$^2$ range. Fig. \ref{expansion_plots}d shows the measured intensity of the 327.4 nm Cu$^0$ and the 404.3 nm Cu$^{+}$ lines on the left axis of the plot. The measured ratio of Cu$^0$:Cu$^{+}$ line intensities is shown on the right axis.  Since reference data for the relative intensities of these lines indicate a 1:2.7 ratio (Cu$^0$:Cu$^{+}$) \cite{NIST_ASD}, the OES data supports a plume dominated by Cu$^0$. This is in agreement with the plume predicted by the simulation at low fluence (Fig. \ref{expansion_plots}a). Also noteworthy is the fact that the ratio of the Cu$^0$ and Cu$^{+}$ lines remains approximately constant for the measured fluence range, implying that the relative densities of the corresponding species do not change appreciably. No Cu$^{2+}$ emission lines were detected for fluences of 1.5-2.7 J/cm$^2$, which is also consistent with the model's prediction for laser fluence below 3 J/cm$^2$ (Fig. \ref{expansion_plots}a). This low fluence regime is widely used for growth of thin films by PLD, especially when layer-by-layer growth is intended.

The presence of species with higher ionization states (2+, 3+, etc.) is commonly reported in the literature for PLD plumes \cite{Kumaki2016,Torrisi2003,Ilyas2011}. However, these reports often refer to higher laser fluences, typically in excess of 5 J/cm$^2$. For laser fluences in the 1-3 J/cm$^2$ OES and our simulation results of Cu ablation jointly support the notion that the plume is dominated by Cu$^0$, with small concentrations of Cu$^{+}$ and electrons at the leading edge of the plume. 

The kinetic energy distribution of ions in the plasma plume is of particular importance for thin film deposition and 2D materials modification. Insight into this distribution can be gained by comparing simulations and measurements of the ion density as a function of the ion time of flight (TOF). This TOF-based analysis is evidently qualitative, since it would be strictly meaningful only for collisionless plasma expansion regimes. Fig. \ref{leading_edge}a shows simulated ion densities at 3.0 cm from the target for Cu ablation with various laser fluences in the 3-7 J/cm$^2$ range. Not surprisingly, the simulation predicts that the overall ion density scales with the laser fluence. The number of fast ions grows with increasing fluence while the peak of the ion density pulse shifts to shorter TOF values. This trend in the simulation is observed in the densities acquired with the planar Langmuir probe placed at 5.5 cm from the target. Fig. \ref{leading_edge}b shows the experimental ion density as a function of TOF obtained via $n_i(t) = I_{ion}(t)/(e \nu A)$, where $I_{ion}(t)$ is the saturation current pulse measured with the planar probe under strong negative bias, $A$ is the probe area, $e$ is elementary charge, and $\nu$ is the plasma flow velocity \cite{Doggett2009}. Beyond the trivial mismatch caused by the different values of $d$, a comparison between absolute numbers for simulated and experimental ion densities illustrates the quantitative limits of the 1D simulation. As discussed in the model description section, the 1D version of the model is not expected to be quantitatively accurate at long distances. The overestimation of the simulated ion density is attributed to the 1D nature of the calculation, which obviously does not allow plasma expansion into the transverse dimensions. This highlights the importance of 3D implementations (e.g., 2D-axisymmetric), which are underway in our group. Simulations at higher dimensionality will undoubtedly reduce the magnitude of the density along the direction normal to the target at long distances.

The ion densities of Figs. \ref{leading_edge}a and \ref{leading_edge}b represent, essentially, the simulated and measured TOF distributions of the ions. It is instructive to convert the TOF to the quantity $\varepsilon = \frac{1}{2} m_{\mathrm{Cu}} \left(\frac{d}{\mathrm{TOF}}\right)^2$, and analyze how the number of ions arriving at distance $d$ from the target is distributed in $\varepsilon$. These converted distributions for simulation and experiment are shown in Figs. \ref{leading_edge}c and \ref{leading_edge}d. The energy $\varepsilon$ does not represent the kinetic energy of the ions, since it does not include the transverse degrees of freedom of their thermal motion and the plasma expansion regime is clearly hydrodynamic rather than collisionless. However, $dN/d\varepsilon$ obtained from $n_i(t)$ by the change of variables \cite{Franghiadakis1999}
\begin{equation}\label{dN_depsilon}
    \frac{dN}{d\varepsilon} = \frac{\nu t^3}{m_{\mathrm{Cu}} d^2} n_i(t),
\end{equation}
where $t\equiv \mathrm{TOF}$, is an approximate representation of the ion kinetic energy distribution. The accuracy of this approximation improves as the plasma flow velocity $\nu$ increases with respect to the thermal velocity of the ions $\nu_{th}$. Indeed, for most of our data $\nu$ is substantially larger than $\nu_{th}$. 

It is apparent in the simulated distributions of Fig. \ref{leading_edge}c, that a dramatic shift in the peak position for $dN/d\varepsilon$ is associated with the plasma shielding threshold. For ablation at 3 J/cm$^2$ (i.e., below shielding threshold), $dN/d\varepsilon$ peaks at $\sim$ 6 eV. On the other hand, the distributions for ablation above the threshold have peaks $\sim$ 22 eV. Morever, the stretching of the high-energy tail of $dN/d\varepsilon$ with increasing laser fluence is consistent with the rising availability of ions traveling with high kinetic energy. The experimental $dN/d\varepsilon$ distributions (Fig. \ref{leading_edge}d) exhibit the same essential trend seen in the simulation, but are marked by a less abrupt transition to the plasma shielding regime. This is due to the lower plasma flow velocities that characterize the experimental plume. Plasma flow velocities are overestimated in the simulation due to the imposed 1D expansion and the higher internal energy of the plasma, caused by the limited energy loss mechanisms included in the model. The lower plasma flow velocities in the experiment are clearly evidenced in the less pronounced high-energy tails of the experimental $dN/d\varepsilon$ curves. The leading edge of the plume is particularly important in thin film deposition and modification, since it contains the most energetic species arriving at the growth substrate. Figs. \ref{leading_edge}c and \ref{leading_edge}d show good correspondence regarding the impact of laser fluence in the generation of high kinetic energy species at the leading edge.    
Additional details about the leading edge of the plume can be inferred from the simulation data shown in Fig. \ref{leading_edge}e. This figure displays the density profiles for neutral and charged species, as well as the vapor temperature at long distances from the target. The simulation outcomes for a fluence below the plasma shielding threshold (3 J/cm$^2$) are compared to the predicted profiles for conditions just above the threshold (4 J/cm$^2$, inset). The calculated temperature at the leading edge of the plume exhibits a well-defined peak of $\sim$ 0.4 eV at 3 J/cm$^2$ and reaching up to $\sim$ 6 eV for 4 J/cm$^2$. These sharp peaks decay to significantly lower, and approximately spatially uniform values, once the highly ionized leading edge of the plume has passed. This is seen in Fig. \ref{leading_edge}e with plasma temperature values shown on the left axes. Fig. \ref{leading_edge}e also plots the neutral, ion, and electron densities on the right axes. In the below-threshold case, singly-charged ions and electrons are concentrated in the hot leading edge where the temperature peaks. In the simulation at 4 J/cm$^2$, Cu$^{2+}$ ions are noted to bunch at the front of the plume, while the Cu$^{+}$ and Cu$^0$ species follow with electrons suitably distributed to preserve the macroscopic neutrality of the plasma. 

The simulation therefore indicates the presence of a hot leading edge of the plume, followed by a much cooler plasma. This high temperature front is consistent with results of other reported simulations \cite{Palya2018, Sibold1991, Garrelie1999, Chen2006, Clair2011}, and is a hydrodynamic effect resulting from the plasma fluid expansion against the simulated vacuum background, which in this case has a number density of 10$^{18}$ atoms/m$^{3}$. The thermal and ionization characteristics of this plasma front, considered together with the more sustained and long-lived profile of the tail of the plume, may be significant in controlling the synthesis of novel materials. This may be particularly relevant for growth and processing of 2D materials in which very low defect concentrations, affected by ionic collisions and/or ultrafast annealing pulses, may represent new parameters for control in the fabrication of these materials.

\section{Conclusion}
\label{S:5}
We have implemented a well-established laser plasma expansion model using a solver with an adaptive Cartesian mesh. The adaptive mesh drastically reduces computational time, allowing for simulation times up to tens of microseconds and distances of centimeters that have not been fully investigated before. Calculations using a 1D version of the implementation were carried out and compared with Langmuir probe and optical emission spectroscopy data acquired during ablation of Cu by a KrF excimer laser. The solver allows extension to 2D-axisymmetric and 3D simulations.    

The simulation describes a well-defined plasma-shielding threshold at fluences of 3-4 J/cm$^2$ for KrF excimer ablation of Cu, in good correspondence with experimental data acquired by planar and cylindrical Langmuir probes. In the range of laser fluences of greatest interest for PLD and 2D materials ($<$ 4 J/cm$^2$), the simulation predicts a Cu ablation plume dominated by Cu$^0$, with small concentrations of Cu$^{+}$ and electrons at the leading edge of the plume. This plume composition is supported by optical emission spectroscopy experiments. For higher laser fluence (e.g., 7 J/cm$^2$), Cu$^{+}$ becomes the dominant plasma species and the leading edge of the plume achieves full ionization. 

Approximate kinetic energy distributions inferred from the simulated ion TOF show the plasma shielding threshold and are consistent with ion TOF measurements by the planar probe. The restricted dimensionality and simplified pathway for collisional/radiative energy loss by the plasma (Bremmstrahlung only), are likely the causes of the more abrupt plasma-shielding transition and higher plasma flow velocities predicted by the simulation. 

The simulations indicate the presence of a high-temperature spike at the front of the plasma expansion. Although characterized by low particle concentrations $\sim$ 10$^{19}$ m$^{-3}$, it is anticipated that in the context of 2D materials processing, low densities of impinging ions with high translational and thermal kinetic energy will play a significant role in processes such as doping, defect annealing, and phase transformations.

\section*{Acknowledgments}
This work was supported in part by the National Science Foundation (NSF) EPSCoR RII-Track-1 Cooperative Agreement OIA-1655280 and by a grant of high performance computing resources and technical support from the Alabama Supercomputer Authority (ASA). SBH acknowledges graduate fellowship support from the National Aeronautics and Space Administration (NASA) Alabama Space Grant Consortium (ASGC) under award NNX15AJ18H. RRA was partially supported by the Department of Energy (DOE) SBIR Project DE-SC0015746. The authors would like to thank Dr. S. Aaron Catledge for assistance with the OES measurements. Any opinions, findings, and conclusions or recommendations expressed in this material are those of the authors and do not necessarily reflect the views of NSF, ASA, NASA or DOE.



\bibliographystyle{apsrev}
\bibliography{main.bib}

\begin{thebibliography}{42}
\expandafter\ifx\csname natexlab\endcsname\relax\def\natexlab#1{#1}\fi
\expandafter\ifx\csname bibnamefont\endcsname\relax
  \def\bibnamefont#1{#1}\fi
\expandafter\ifx\csname bibfnamefont\endcsname\relax
  \def\bibfnamefont#1{#1}\fi
\expandafter\ifx\csname citenamefont\endcsname\relax
  \def\citenamefont#1{#1}\fi
\expandafter\ifx\csname url\endcsname\relax
  \def\url#1{\texttt{#1}}\fi
\expandafter\ifx\csname urlprefix\endcsname\relax\def\urlprefix{URL }\fi
\providecommand{\bibinfo}[2]{#2}
\providecommand{\eprint}[2][]{\url{#2}}

\bibitem[{\citenamefont{Chang and Chang}(2017)}]{Chang2017}
\bibinfo{author}{\bibfnamefont{J.}~\bibnamefont{Chang}} \bibnamefont{and}
  \bibinfo{author}{\bibfnamefont{J.~P.} \bibnamefont{Chang}},
  \bibinfo{journal}{J Phys. D Appl. Phys.} \textbf{\bibinfo{volume}{50}},
  \bibinfo{pages}{253001} (\bibinfo{year}{2017}).

\bibitem[{\citenamefont{Allain and Shetty}(2017)}]{Allain2017}
\bibinfo{author}{\bibfnamefont{J.~P.} \bibnamefont{Allain}} \bibnamefont{and}
  \bibinfo{author}{\bibfnamefont{A.}~\bibnamefont{Shetty}}, \bibinfo{journal}{J
  Phys. D Appl. Phys.} \textbf{\bibinfo{volume}{50}}, \bibinfo{pages}{283002}
  (\bibinfo{year}{2017}).

\bibitem[{\citenamefont{Oehrlein and Hamaguchi}(2018)}]{Oehrlein2018}
\bibinfo{author}{\bibfnamefont{G.~S.} \bibnamefont{Oehrlein}} \bibnamefont{and}
  \bibinfo{author}{\bibfnamefont{S.}~\bibnamefont{Hamaguchi}},
  \bibinfo{journal}{Plasma Sources Sci. Technol.}
  \textbf{\bibinfo{volume}{27}}, \bibinfo{pages}{023001}
  (\bibinfo{year}{2018}).

\bibitem[{\citenamefont{Balasubramanyam
  et~al.}(2019)\citenamefont{Balasubramanyam, Shirazi, Bloodgood, Longfei,
  Verheijen, Vandalon, Kessels, Hofmann, and Bol}}]{Balasubramanyam2019a}
\bibinfo{author}{\bibfnamefont{S.}~\bibnamefont{Balasubramanyam}},
  \bibinfo{author}{\bibfnamefont{M.}~\bibnamefont{Shirazi}},
  \bibinfo{author}{\bibfnamefont{M.}~\bibnamefont{Bloodgood}},
  \bibinfo{author}{\bibfnamefont{W.}~\bibnamefont{Longfei}},
  \bibinfo{author}{\bibfnamefont{M.}~\bibnamefont{Verheijen}},
  \bibinfo{author}{\bibfnamefont{V.}~\bibnamefont{Vandalon}},
  \bibinfo{author}{\bibfnamefont{W.}~\bibnamefont{Kessels}},
  \bibinfo{author}{\bibfnamefont{J.}~\bibnamefont{Hofmann}}, \bibnamefont{and}
  \bibinfo{author}{\bibfnamefont{A.}~\bibnamefont{Bol}},
  \bibinfo{journal}{Chem. Mater.} \textbf{\bibinfo{volume}{31}},
  \bibinfo{pages}{5104} (\bibinfo{year}{2019}).

\bibitem[{\citenamefont{Oyedele et~al.}(2019)\citenamefont{Oyedele, Yang, Feng,
  Haglund, Gu, Puretzky, Briggs, Rouleau, Chisholm, Unocic
  et~al.}}]{Oyedele2019}
\bibinfo{author}{\bibfnamefont{A.}~\bibnamefont{Oyedele}},
  \bibinfo{author}{\bibfnamefont{S.}~\bibnamefont{Yang}},
  \bibinfo{author}{\bibfnamefont{T.}~\bibnamefont{Feng}},
  \bibinfo{author}{\bibfnamefont{A.}~\bibnamefont{Haglund}},
  \bibinfo{author}{\bibfnamefont{Y.}~\bibnamefont{Gu}},
  \bibinfo{author}{\bibfnamefont{A.}~\bibnamefont{Puretzky}},
  \bibinfo{author}{\bibfnamefont{D.}~\bibnamefont{Briggs}},
  \bibinfo{author}{\bibfnamefont{C.}~\bibnamefont{Rouleau}},
  \bibinfo{author}{\bibfnamefont{M.}~\bibnamefont{Chisholm}},
  \bibinfo{author}{\bibfnamefont{R.}~\bibnamefont{Unocic}},
  \bibnamefont{et~al.}, \bibinfo{journal}{J. Am. Chem. Soc.}
  \textbf{\bibinfo{volume}{141}}, \bibinfo{pages}{8928} (\bibinfo{year}{2019}).

\bibitem[{\citenamefont{Harilal et~al.}(2016)\citenamefont{Harilal, Brumfield,
  Cannon, and Phillips}}]{Harilal2016}
\bibinfo{author}{\bibfnamefont{S.~S.} \bibnamefont{Harilal}},
  \bibinfo{author}{\bibfnamefont{B.~E.} \bibnamefont{Brumfield}},
  \bibinfo{author}{\bibfnamefont{B.~D.} \bibnamefont{Cannon}},
  \bibnamefont{and} \bibinfo{author}{\bibfnamefont{M.~C.}
  \bibnamefont{Phillips}}, \bibinfo{journal}{Anal. Chem.}
  \textbf{\bibinfo{volume}{88}}, \bibinfo{pages}{2296} (\bibinfo{year}{2016}).

\bibitem[{\citenamefont{Liu et~al.}(2019)\citenamefont{Liu, Ashfold, Meehan,
  and Wagenaars}}]{Liu2019}
\bibinfo{author}{\bibfnamefont{H.}~\bibnamefont{Liu}},
  \bibinfo{author}{\bibfnamefont{M.~N.} \bibnamefont{Ashfold}},
  \bibinfo{author}{\bibfnamefont{D.~N.} \bibnamefont{Meehan}},
  \bibnamefont{and}
  \bibinfo{author}{\bibfnamefont{E.}~\bibnamefont{Wagenaars}},
  \bibinfo{journal}{J. Appl. Phys.} \textbf{\bibinfo{volume}{125}},
  \bibinfo{pages}{083304} (\bibinfo{year}{2019}).

\bibitem[{\citenamefont{Liu et~al.}(2016)\citenamefont{Liu, Truscott, and
  Ashfold}}]{Liu2016}
\bibinfo{author}{\bibfnamefont{H.}~\bibnamefont{Liu}},
  \bibinfo{author}{\bibfnamefont{B.~S.} \bibnamefont{Truscott}},
  \bibnamefont{and} \bibinfo{author}{\bibfnamefont{M.~N.~R.}
  \bibnamefont{Ashfold}}, \bibinfo{journal}{Sci. Rep.}
  \textbf{\bibinfo{volume}{6}}, \bibinfo{pages}{25609} (\bibinfo{year}{2016}).

\bibitem[{\citenamefont{Geohegan}(1992)}]{Geohegan1992a}
\bibinfo{author}{\bibfnamefont{D.}~\bibnamefont{Geohegan}}, in
  \emph{\bibinfo{booktitle}{Laser Ablation of Electronic Materials: Basic
  Mechanisms and Applications}}, edited by
  \bibinfo{editor}{\bibfnamefont{E.}~\bibnamefont{Fogarassy}} \bibnamefont{and}
  \bibinfo{editor}{\bibfnamefont{S.}~\bibnamefont{Lazare}}
  (\bibinfo{publisher}{New Holland}, \bibinfo{address}{Amsterdam},
  \bibinfo{year}{1992}), pp. \bibinfo{pages}{73--88}.

\bibitem[{\citenamefont{Willmott and Huber}(2000)}]{Willmott2000}
\bibinfo{author}{\bibfnamefont{P.~R.} \bibnamefont{Willmott}} \bibnamefont{and}
  \bibinfo{author}{\bibfnamefont{J.~R.} \bibnamefont{Huber}},
  \bibinfo{journal}{Rev. Mod. Phys.} \textbf{\bibinfo{volume}{72}},
  \bibinfo{pages}{315} (\bibinfo{year}{2000}).

\bibitem[{\citenamefont{Geohegan}(1994)}]{Geohegan1994}
\bibinfo{author}{\bibfnamefont{D.}~\bibnamefont{Geohegan}}, in
  \emph{\bibinfo{booktitle}{Pulsed Laser Deposition of Thin Films}}, edited by
  \bibinfo{editor}{\bibfnamefont{D.}~\bibnamefont{Chrisey}} \bibnamefont{and}
  \bibinfo{editor}{\bibfnamefont{G.}~\bibnamefont{Hubler}}
  (\bibinfo{publisher}{Wiley-Interscience}, \bibinfo{address}{New York},
  \bibinfo{year}{1994}), pp. \bibinfo{pages}{115--163}.

\bibitem[{\citenamefont{Palya et~al.}(2018)\citenamefont{Palya, Ranjbar, Lin,
  and Volkov}}]{Palya2018}
\bibinfo{author}{\bibfnamefont{A.}~\bibnamefont{Palya}},
  \bibinfo{author}{\bibfnamefont{O.~A.} \bibnamefont{Ranjbar}},
  \bibinfo{author}{\bibfnamefont{Z.}~\bibnamefont{Lin}}, \bibnamefont{and}
  \bibinfo{author}{\bibfnamefont{A.~N.} \bibnamefont{Volkov}},
  \bibinfo{journal}{Appl. Phys. A Mater. Sci. Process.}
  \textbf{\bibinfo{volume}{124}}, \bibinfo{pages}{32} (\bibinfo{year}{2018}).

\bibitem[{\citenamefont{Shabanov and Gornushkin}(2014)}]{Shabanov2014}
\bibinfo{author}{\bibfnamefont{S.}~\bibnamefont{Shabanov}} \bibnamefont{and}
  \bibinfo{author}{\bibfnamefont{I.}~\bibnamefont{Gornushkin}},
  \bibinfo{journal}{Spectrochim. Acta B} \textbf{\bibinfo{volume}{100}},
  \bibinfo{pages}{147} (\bibinfo{year}{2014}).

\bibitem[{\citenamefont{Cadot et~al.}(2018)\citenamefont{Cadot, Thomas, Best,
  Taylor, Michler, Axinte, and Billingham}}]{Cadot2018}
\bibinfo{author}{\bibfnamefont{G.}~\bibnamefont{Cadot}},
  \bibinfo{author}{\bibfnamefont{K.}~\bibnamefont{Thomas}},
  \bibinfo{author}{\bibfnamefont{J.}~\bibnamefont{Best}},
  \bibinfo{author}{\bibfnamefont{A.}~\bibnamefont{Taylor}},
  \bibinfo{author}{\bibfnamefont{J.}~\bibnamefont{Michler}},
  \bibinfo{author}{\bibfnamefont{D.}~\bibnamefont{Axinte}}, \bibnamefont{and}
  \bibinfo{author}{\bibfnamefont{J.}~\bibnamefont{Billingham}},
  \bibinfo{journal}{Carbon} \textbf{\bibinfo{volume}{127}},
  \bibinfo{pages}{349} (\bibinfo{year}{2018}).

\bibitem[{\citenamefont{Pathak and Povitsky}(2008)}]{Pathak2008}
\bibinfo{author}{\bibfnamefont{K.}~\bibnamefont{Pathak}} \bibnamefont{and}
  \bibinfo{author}{\bibfnamefont{A.}~\bibnamefont{Povitsky}},
  \bibinfo{journal}{J. Appl. Phys.} \textbf{\bibinfo{volume}{104}},
  \bibinfo{pages}{113108} (\bibinfo{year}{2008}).

\bibitem[{\citenamefont{Sizyuk and Hassanein}(2014)}]{Sizyuk2014}
\bibinfo{author}{\bibfnamefont{T.}~\bibnamefont{Sizyuk}} \bibnamefont{and}
  \bibinfo{author}{\bibfnamefont{A.}~\bibnamefont{Hassanein}},
  \bibinfo{journal}{Nuc. Fusion} \textbf{\bibinfo{volume}{54}},
  \bibinfo{pages}{023004} (\bibinfo{year}{2014}).

\bibitem[{\citenamefont{Yadav et~al.}(2017)\citenamefont{Yadav, Patel, Singh,
  Das, Kaw, and Kumar}}]{Sharad2017}
\bibinfo{author}{\bibfnamefont{Y.}~\bibnamefont{Yadav}},
  \bibinfo{author}{\bibfnamefont{B.}~\bibnamefont{Patel}},
  \bibinfo{author}{\bibfnamefont{R.}~\bibnamefont{Singh}},
  \bibinfo{author}{\bibfnamefont{A.}~\bibnamefont{Das}},
  \bibinfo{author}{\bibfnamefont{P.}~\bibnamefont{Kaw}}, \bibnamefont{and}
  \bibinfo{author}{\bibfnamefont{A.}~\bibnamefont{Kumar}}, \bibinfo{journal}{J.
  Phys. D. Appl. Phys.} \textbf{\bibinfo{volume}{50}}, \bibinfo{pages}{355201}
  (\bibinfo{year}{2017}).

\bibitem[{\citenamefont{Pietanza et~al.}(2010)\citenamefont{Pietanza, Colonna,
  De~Giacomo, and Capitelli}}]{Pietanza2010}
\bibinfo{author}{\bibfnamefont{L.}~\bibnamefont{Pietanza}},
  \bibinfo{author}{\bibfnamefont{G.}~\bibnamefont{Colonna}},
  \bibinfo{author}{\bibfnamefont{A.}~\bibnamefont{De~Giacomo}},
  \bibnamefont{and}
  \bibinfo{author}{\bibfnamefont{M.}~\bibnamefont{Capitelli}},
  \bibinfo{journal}{Spectrochim. Acta B} \textbf{\bibinfo{volume}{65}},
  \bibinfo{pages}{616} (\bibinfo{year}{2010}).

\bibitem[{\citenamefont{Chen and Bogaerts}(2005)}]{Chen2005}
\bibinfo{author}{\bibfnamefont{Z.}~\bibnamefont{Chen}} \bibnamefont{and}
  \bibinfo{author}{\bibfnamefont{A.}~\bibnamefont{Bogaerts}},
  \bibinfo{journal}{J. Appl. Phys.} \textbf{\bibinfo{volume}{97}},
  \bibinfo{pages}{063305} (\bibinfo{year}{2005}).

\bibitem[{\citenamefont{Kelly and Miotello}(1994)}]{KellyandMiotello}
\bibinfo{author}{\bibfnamefont{R.}~\bibnamefont{Kelly}} \bibnamefont{and}
  \bibinfo{author}{\bibfnamefont{A.}~\bibnamefont{Miotello}}, in
  \emph{\bibinfo{booktitle}{Pulsed Laser Deposition of Thin Films}}, edited by
  \bibinfo{editor}{\bibfnamefont{D.}~\bibnamefont{Chrisey}} \bibnamefont{and}
  \bibinfo{editor}{\bibfnamefont{G.}~\bibnamefont{Hubler}}
  (\bibinfo{publisher}{Wiley-Interscience}, \bibinfo{address}{New York},
  \bibinfo{year}{1994}), pp. \bibinfo{pages}{55--87}.

\bibitem[{\citenamefont{Autrique
  et~al.}(2013{\natexlab{a}})\citenamefont{Autrique, Clair, L'Hermite,
  Alexiades, Bogaerts, and Rethfeld}}]{Autrique2013}
\bibinfo{author}{\bibfnamefont{D.}~\bibnamefont{Autrique}},
  \bibinfo{author}{\bibfnamefont{G.}~\bibnamefont{Clair}},
  \bibinfo{author}{\bibfnamefont{D.}~\bibnamefont{L'Hermite}},
  \bibinfo{author}{\bibfnamefont{V.}~\bibnamefont{Alexiades}},
  \bibinfo{author}{\bibfnamefont{A.}~\bibnamefont{Bogaerts}}, \bibnamefont{and}
  \bibinfo{author}{\bibfnamefont{B.}~\bibnamefont{Rethfeld}},
  \bibinfo{journal}{J. Appl. Phys.} \textbf{\bibinfo{volume}{114}},
  \bibinfo{pages}{023301} (\bibinfo{year}{2013}{\natexlab{a}}).

\bibitem[{\citenamefont{Autrique
  et~al.}(2013{\natexlab{b}})\citenamefont{Autrique, Gornushkin, Alexiades,
  Chen, Bogaerts, and Rethfeld}}]{Autrique2013a}
\bibinfo{author}{\bibfnamefont{D.}~\bibnamefont{Autrique}},
  \bibinfo{author}{\bibfnamefont{I.}~\bibnamefont{Gornushkin}},
  \bibinfo{author}{\bibfnamefont{V.}~\bibnamefont{Alexiades}},
  \bibinfo{author}{\bibfnamefont{Z.}~\bibnamefont{Chen}},
  \bibinfo{author}{\bibfnamefont{A.}~\bibnamefont{Bogaerts}}, \bibnamefont{and}
  \bibinfo{author}{\bibfnamefont{B.}~\bibnamefont{Rethfeld}},
  \bibinfo{journal}{Appl. Phys. Lett.} \textbf{\bibinfo{volume}{103}},
  \bibinfo{pages}{174102} (\bibinfo{year}{2013}{\natexlab{b}}).

\bibitem[{\citenamefont{Palya et~al.}(2019)\citenamefont{Palya, Ranjbar, Lin,
  and Volkov}}]{Palya2019}
\bibinfo{author}{\bibfnamefont{A.}~\bibnamefont{Palya}},
  \bibinfo{author}{\bibfnamefont{O.~A.} \bibnamefont{Ranjbar}},
  \bibinfo{author}{\bibfnamefont{Z.}~\bibnamefont{Lin}}, \bibnamefont{and}
  \bibinfo{author}{\bibfnamefont{A.~N.} \bibnamefont{Volkov}},
  \bibinfo{journal}{Int. J. Heat Mass Transf.} p. \bibinfo{pages}{1029}
  (\bibinfo{year}{2019}).

\bibitem[{\citenamefont{Wang et~al.}(2016)\citenamefont{Wang, Yuan, Fu, and
  Wang}}]{Wang2016}
\bibinfo{author}{\bibfnamefont{Y.}~\bibnamefont{Wang}},
  \bibinfo{author}{\bibfnamefont{H.}~\bibnamefont{Yuan}},
  \bibinfo{author}{\bibfnamefont{Y.}~\bibnamefont{Fu}}, \bibnamefont{and}
  \bibinfo{author}{\bibfnamefont{Z.}~\bibnamefont{Wang}},
  \bibinfo{journal}{Spectrochim. Acta B} \textbf{\bibinfo{volume}{126}},
  \bibinfo{pages}{44} (\bibinfo{year}{2016}).

\bibitem[{\citenamefont{Oumeziane et~al.}(2016)\citenamefont{Oumeziane, Liani,
  and Parisse}}]{Oumeziane2016}
\bibinfo{author}{\bibfnamefont{A.~A.} \bibnamefont{Oumeziane}},
  \bibinfo{author}{\bibfnamefont{B.}~\bibnamefont{Liani}}, \bibnamefont{and}
  \bibinfo{author}{\bibfnamefont{J.~D.} \bibnamefont{Parisse}},
  \bibinfo{journal}{Plasma Chem. Plasma Process.}
  \textbf{\bibinfo{volume}{36}}, \bibinfo{pages}{711} (\bibinfo{year}{2016}).

\bibitem[{\citenamefont{Abdellatif and Imam}(2002)}]{Abdellatif2002}
\bibinfo{author}{\bibfnamefont{G.}~\bibnamefont{Abdellatif}} \bibnamefont{and}
  \bibinfo{author}{\bibfnamefont{H.}~\bibnamefont{Imam}},
  \bibinfo{journal}{Spectrochim. Acta B} \textbf{\bibinfo{volume}{57}},
  \bibinfo{pages}{1155} (\bibinfo{year}{2002}).

\bibitem[{bas()}]{basilisk.fr}
\bibinfo{note}{See http://basilisk.fr/ for Basilisk: a Free Software program
  for the solution of partial differential equations on adaptive Cartesian
  meshes.}

\bibitem[{\citenamefont{Spitzer}(2006)}]{Spitzer2006}
\bibinfo{author}{\bibfnamefont{L.}~\bibnamefont{Spitzer}},
  \emph{\bibinfo{title}{Physics of Fully Ionized Gases}}
  (\bibinfo{publisher}{Dover Publications, Inc.}, \bibinfo{address}{Mineola},
  \bibinfo{year}{2006}), p. \bibinfo{pages}{148}.

\bibitem[{\citenamefont{Zel'dovich and Raizer}(2002)}]{Zeldovich2002}
\bibinfo{author}{\bibfnamefont{Y.}~\bibnamefont{Zel'dovich}} \bibnamefont{and}
  \bibinfo{author}{\bibfnamefont{Y.}~\bibnamefont{Raizer}},
  \emph{\bibinfo{title}{Physics of Shock Waves and High-Temperature
  Hydrodynamic Phenomena}} (\bibinfo{publisher}{Dover},
  \bibinfo{address}{Mineola}, \bibinfo{year}{2002}), pp.
  \bibinfo{pages}{192--195}.

\bibitem[{\citenamefont{Root}(1989)}]{Rakziemski1989}
\bibinfo{author}{\bibfnamefont{R.}~\bibnamefont{Root}}, in
  \emph{\bibinfo{booktitle}{Laser-Induced Plasmas and Applications}}, edited by
  \bibinfo{editor}{\bibfnamefont{L.}~\bibnamefont{Radziemski}}
  \bibnamefont{and} \bibinfo{editor}{\bibfnamefont{D.}~\bibnamefont{Cremers}}
  (\bibinfo{publisher}{Marcel Dekker, Inc.}, \bibinfo{address}{New York},
  \bibinfo{year}{1989}), pp. \bibinfo{pages}{69--103}.

\bibitem[{\citenamefont{Doggett and Lunney}(2009)}]{Doggett2009}
\bibinfo{author}{\bibfnamefont{B.}~\bibnamefont{Doggett}} \bibnamefont{and}
  \bibinfo{author}{\bibfnamefont{J.}~\bibnamefont{Lunney}},
  \bibinfo{journal}{J. Appl. Phys.} \textbf{\bibinfo{volume}{105}},
  \bibinfo{pages}{033306} (\bibinfo{year}{2009}).

\bibitem[{\citenamefont{Weaver et~al.}(1999)\citenamefont{Weaver, Martin,
  Graham, Morrow, and Lewis}}]{Weaver1999}
\bibinfo{author}{\bibfnamefont{I.}~\bibnamefont{Weaver}},
  \bibinfo{author}{\bibfnamefont{G.}~\bibnamefont{Martin}},
  \bibinfo{author}{\bibfnamefont{W.}~\bibnamefont{Graham}},
  \bibinfo{author}{\bibfnamefont{T.}~\bibnamefont{Morrow}}, \bibnamefont{and}
  \bibinfo{author}{\bibfnamefont{C.}~\bibnamefont{Lewis}},
  \bibinfo{journal}{Rev. Sci. Instrum.} \textbf{\bibinfo{volume}{70}},
  \bibinfo{pages}{1801} (\bibinfo{year}{1999}).

\bibitem[{\citenamefont{Mao and Russo}(1997)}]{Mao1997}
\bibinfo{author}{\bibfnamefont{X.}~\bibnamefont{Mao}} \bibnamefont{and}
  \bibinfo{author}{\bibfnamefont{R.}~\bibnamefont{Russo}},
  \bibinfo{journal}{Appl. Phys. A} \textbf{\bibinfo{volume}{64}},
  \bibinfo{pages}{1} (\bibinfo{year}{1997}).

\bibitem[{\citenamefont{Kramida et~al.}(2018)\citenamefont{Kramida,
  {Yu.~Ralchenko}, Reader, and {and NIST ASD Team}}}]{NIST_ASD}
\bibinfo{author}{\bibfnamefont{A.}~\bibnamefont{Kramida}},
  \bibinfo{author}{\bibnamefont{{Yu.~Ralchenko}}},
  \bibinfo{author}{\bibfnamefont{J.}~\bibnamefont{Reader}}, \bibnamefont{and}
  \bibinfo{author}{\bibnamefont{{and NIST ASD Team}}},
  \bibinfo{howpublished}{{NIST Atomic Spectra Database (ver. 5.6.1), [Online].
  Available: {\tt{https://physics.nist.gov/asd}} [2019, August 8]. National
  Institute of Standards and Technology, Gaithersburg, MD.}}
  (\bibinfo{year}{2018}).

\bibitem[{\citenamefont{Kumaki et~al.}(2016)\citenamefont{Kumaki, Steski,
  Ikeda, Kanesue, Okamura, and Washio}}]{Kumaki2016}
\bibinfo{author}{\bibfnamefont{M.}~\bibnamefont{Kumaki}},
  \bibinfo{author}{\bibfnamefont{D.}~\bibnamefont{Steski}},
  \bibinfo{author}{\bibfnamefont{S.}~\bibnamefont{Ikeda}},
  \bibinfo{author}{\bibfnamefont{T.}~\bibnamefont{Kanesue}},
  \bibinfo{author}{\bibfnamefont{M.}~\bibnamefont{Okamura}}, \bibnamefont{and}
  \bibinfo{author}{\bibfnamefont{M.}~\bibnamefont{Washio}},
  \bibinfo{journal}{Rev. Sci. Instrum.} \textbf{\bibinfo{volume}{87}},
  \bibinfo{pages}{02A921} (\bibinfo{year}{2016}).

\bibitem[{\citenamefont{Torrisi et~al.}(2003)\citenamefont{Torrisi, Gammino,
  Ando, Nassisi, Doria, and Perdone}}]{Torrisi2003}
\bibinfo{author}{\bibfnamefont{L.}~\bibnamefont{Torrisi}},
  \bibinfo{author}{\bibfnamefont{S.}~\bibnamefont{Gammino}},
  \bibinfo{author}{\bibfnamefont{L.}~\bibnamefont{Ando}},
  \bibinfo{author}{\bibfnamefont{V.}~\bibnamefont{Nassisi}},
  \bibinfo{author}{\bibfnamefont{D.}~\bibnamefont{Doria}}, \bibnamefont{and}
  \bibinfo{author}{\bibfnamefont{A.}~\bibnamefont{Perdone}},
  \bibinfo{journal}{Appl. Surf. Sci.} \textbf{\bibinfo{volume}{210}},
  \bibinfo{pages}{262} (\bibinfo{year}{2003}).

\bibitem[{\citenamefont{Ilyas et~al.}(2001)\citenamefont{Ilyas, Dogar, Ullah,
  and Qayyum}}]{Ilyas2011}
\bibinfo{author}{\bibfnamefont{B.}~\bibnamefont{Ilyas}},
  \bibinfo{author}{\bibfnamefont{A.}~\bibnamefont{Dogar}},
  \bibinfo{author}{\bibfnamefont{S.}~\bibnamefont{Ullah}}, \bibnamefont{and}
  \bibinfo{author}{\bibfnamefont{A.}~\bibnamefont{Qayyum}},
  \bibinfo{journal}{J. Phys. D: Appl. Phys.} \textbf{\bibinfo{volume}{44}},
  \bibinfo{pages}{295202} (\bibinfo{year}{2001}).

\bibitem[{\citenamefont{Franghiadakis et~al.}(1999)\citenamefont{Franghiadakis,
  Fotakis, and Tzanetakis}}]{Franghiadakis1999}
\bibinfo{author}{\bibfnamefont{Y.}~\bibnamefont{Franghiadakis}},
  \bibinfo{author}{\bibfnamefont{C.}~\bibnamefont{Fotakis}}, \bibnamefont{and}
  \bibinfo{author}{\bibfnamefont{P.}~\bibnamefont{Tzanetakis}},
  \bibinfo{journal}{Appl. Phys. A} \textbf{\bibinfo{volume}{68}},
  \bibinfo{pages}{391} (\bibinfo{year}{1999}).

\bibitem[{\citenamefont{Sibold and Urbassek}(1991)}]{Sibold1991}
\bibinfo{author}{\bibfnamefont{D.}~\bibnamefont{Sibold}} \bibnamefont{and}
  \bibinfo{author}{\bibfnamefont{H.}~\bibnamefont{Urbassek}},
  \bibinfo{journal}{Phys. Rev. A} \textbf{\bibinfo{volume}{43}},
  \bibinfo{pages}{6722} (\bibinfo{year}{1991}).

\bibitem[{\citenamefont{Garrelie et~al.}(1999)\citenamefont{Garrelie,
  Champeaux, and Catherinot}}]{Garrelie1999}
\bibinfo{author}{\bibfnamefont{F.}~\bibnamefont{Garrelie}},
  \bibinfo{author}{\bibfnamefont{C.}~\bibnamefont{Champeaux}},
  \bibnamefont{and}
  \bibinfo{author}{\bibfnamefont{A.}~\bibnamefont{Catherinot}},
  \bibinfo{journal}{Appl. Phys. A} \textbf{\bibinfo{volume}{69}},
  \bibinfo{pages}{45} (\bibinfo{year}{1999}).

\bibitem[{\citenamefont{Chen et~al.}(2006)\citenamefont{Chen, Bleiner, and
  Bogaerts}}]{Chen2006}
\bibinfo{author}{\bibfnamefont{Z.}~\bibnamefont{Chen}},
  \bibinfo{author}{\bibfnamefont{D.}~\bibnamefont{Bleiner}}, \bibnamefont{and}
  \bibinfo{author}{\bibfnamefont{A.}~\bibnamefont{Bogaerts}},
  \bibinfo{journal}{J. Appl. Phys.} \textbf{\bibinfo{volume}{99}},
  \bibinfo{pages}{063304} (\bibinfo{year}{2006}).

\bibitem[{\citenamefont{Clair and L'Hermite}(2011)}]{Clair2011}
\bibinfo{author}{\bibfnamefont{G.}~\bibnamefont{Clair}} \bibnamefont{and}
  \bibinfo{author}{\bibfnamefont{D.}~\bibnamefont{L'Hermite}},
  \bibinfo{journal}{J. Appl. Phys.} \textbf{\bibinfo{volume}{110}},
  \bibinfo{pages}{083307} (\bibinfo{year}{2011}).

\end{thebibliography}

\end{document}